\documentclass[lettersize,journal]{IEEEtran}
\usepackage{amsmath,amsfonts}
\usepackage{algorithmic}
\usepackage{algorithm}
\usepackage{array}
\usepackage[caption=false,font=normalsize,labelfont=sf,textfont=sf]{subfig}
\usepackage{textcomp}
\usepackage{balance}
\usepackage{stfloats}
\usepackage{url}
\usepackage{verbatim}
\usepackage{soul}
\usepackage{xcolor}
\usepackage{graphicx}
\usepackage{cite}
\usepackage{float}
\usepackage{placeins}
\usepackage{accessibility}
\newcounter{panel}

\usepackage{adjustbox}
\usepackage{xurl}
\AtBeginDocument{%
  }

\begin{document}
\title{Privacy-Preserving Clustering Algorithm and Parameter Selection\\ for Collaborative Population Stratification \\Using Distributed PCA}

\author{Maryam Ghasemian, Lynette Hammond Gerido, Erman Ayday%
\thanks{Maryam Ghasemian and Erman Ayday are with the Department of Computer and Data Sciences , Case School of Engineering, Case Western Reserve University, Cleveland, OH, USA. Email: \{maryam.ghasemian, erman.ayday\}@case.edu}%
\thanks{Lynette Hammond Gerido is with the Department of Bioethics, School of Medicine, Case Western Reserve University, Cleveland, OH, USA. Email: LHGerido@case.edu}}

\maketitle

\begin{abstract}
We propose a privacy-preserving selection layer for collaborative population stratification workflows, enabling an \textit{honest-but-curious server} to recommend a strong clustering algorithm-hyperparameter configuration from privacy protected shares. Instead of using a fixed workflow such as PCA followed by K-Means with a preset number of clusters, the framework compares multiple clustering methods and parameter settings under privacy constraints. To avoid relying on a public reference basis, we construct a shared low-dimensional representation through a distributed-PCA (DPCA) procedure in which each party computes local top-$r$ principal-component information and the server aggregates only low-rank summaries to recover a global basis. Building on this representation, the framework supports three LDP-protected pipelines: PCA then Noise, Noise then PCA, and Noise-Only. The server evaluates clustering methods and parameter settings using internal validation metrics and recommends a top-performing configuration from the perturbed data; this step is pure post-processing.

Experiments on openSNP and 1000 Genomes show that the server-side selection layer reliably identifies strong settings from LDP-protected data. Across both datasets, the PCA-based pipelines outperform \textit{Noise-Only} in clustering utility, communication cost, and runtime, and show lower membership-inference attack (MIA) power at small privacy budgets. Overall, they provide the best utility--scalability trade-off in our setting.

\end{abstract}

  
\section{Introduction}\label{sec:introduction}
Population stratification (i.e., the existence of subpopulation structure within genetic data) can distort downstream analyses such as association studies and clustering-based phenotype prediction. In collaborative scenarios, multiple parties (researchers) each possess high-dimensional genotype or phenotype datasets and aim to jointly stratify the combined population without exposing sensitive individual‐level information. 
Prior privacy-preserving work has shown that secure multi-party computation (MPC) and local differential privacy (LDP) can make collaborative clustering feasible. However, nearly all existing protocols commit to a single clustering pipeline and fixed hyperparameters—most often principal component analysis (PCA) followed by K-Means with a predetermined number of clusters $K$. Because clustering quality is highly dependent on the data, this “one-size-fits-all” design can yield suboptimal subgroup discovery. A framework that can decide, based on noisy data, which algorithm and parameter settings to use would bridge this configuration gap.

In this paper, we propose a privacy-preserving selection layer for collaborative population stratification. The goal is not to introduce a new clustering method for stratification itself, but to help collaborating parties choose an effective clustering algorithm–parameter pair under privacy constraints. In particular, rather than assuming that PCA followed by K-Means with a preset $K$ is always appropriate, the \textit{honest-but-curious server} evaluates candidate configurations on privacy-protected shares and recommends a strong configuration for the dataset at hand. Also, to accommodate the diverse resources and preferences of collaborating parties (researchers), we introduce three $\epsilon$-LDP pipelines, each representing a different trade-off among researcher-side and server-side computational costs, utility, and privacy. 

We employ a distributed-PCA (DPCA) framework followed by server-side algorithm and hyperparameter selection:

\begin{enumerate}\label{DP_pipeline}
  \item \textbf{PCA\,$\rightarrow$\,Noise}: Each party projects its data onto $r\ll d$ local principal components and then adds Laplace noise to the resulting coordinates before sharing them with the server. This requires relatively modest local computation and reduces communication cost.
  \item \textbf{Noise\,$\rightarrow$\,PCA}: Each party first injects Laplace noise into the raw data and then runs PCA on the noisy data before sharing the projected outputs with the server. The initial perturbation immediately ensures $\epsilon$-LDP, while the subsequent projection reduces the size of the shared data without incurring additional privacy loss.
  \item \textbf{Noise-Only}: Each party adds Laplace noise directly to the full $d$-dimensional raw data and shares the entire noisy dataset with the server, without any dimensionality reduction (thus avoiding PCA overhead).
\end{enumerate}

Because none of these pipelines uniformly dominates the others across all conditions, we treat them as complementary choices that collaborators can select based on their hardware constraints, desired accuracy, and privacy budgets.

After receiving the $\epsilon$-LDP shares, the honest-but-curious server explores a set of candidate clustering configurations, including K-Means, Gaussian Mixture Models (GMM), Agglomerative Clustering (HC), and Density-Based Spatial Clustering of Applications with Noise (DBSCAN), over a range of hyperparameter settings. Each configuration is evaluated using a weighted combination of the silhouette score, Calinski--Harabasz index, and Davies--Bouldin index, and the highest-scoring configuration is reported back to the parties. Because this search is performed entirely on perturbed data, it incurs no additional privacy loss beyond the local $\epsilon$-LDP perturbation step.

Using real genomic datasets, we show that: (1) the server-side selection layer consistently identifies one of the top-performing algorithm-hyperparameter settings from LDP-protected data; (2) the PCA-based pipelines provide substantially better clustering utility than \textit{Noise-Only}; (3) under the distributed-PCA framework, the PCA-based pipelines exhibit lower MIA power in the low-$\epsilon$ regime, whereas \textit{Noise-Only} reaches high attack power more quickly; and (4) the PCA-based pipelines drastically reduce communication and runtime by avoiding clustering in the original high-dimensional SNP space.

A preliminary version of this work appeared in \cite{Ghasemian2025PrivacyCollaborative}. The present version extends that paper by replacing the public PCA model with a distributed-PCA construction introduced in \cite{Fan2019DistributedPCA}, enlarging the clustering search space to include DBSCAN, and expanding the evaluation from the earlier public-PCA/openSNP setting to the broader distributed-PCA experiments on openSNP and 1000 Genomes.
Our main contributions can be summarized as follows:
\begin{itemize}
    \item We design a privacy-preserving selection layer for collaborative population stratification that combines distributed-PCA with a privacy-preserving server-side selection layer for clustering under $\epsilon$-LDP and recommends a strong clustering algorithm-hyperparameter pair, instead of relying on commonly used fixed workflows such as PCA + K-Means with preset $K$.
    \item We support and systematically compare three complementary pipelines---\textit{PCA$\rightarrow$Noise}, \textit{Noise$\rightarrow$PCA}, and \textit{Noise-Only}---under the same privacy budget.
    \item We extend the clustering search space 
    by incorporating \textbf{DBSCAN} alongside K-Means, GMM, and HC, and we perform automated algorithm and hyperparameter selection using only aggregated LDP-protected data.
    \item We provide an empirical evaluation on openSNP and 1000 Genomes datasets showing that the PCA-based pipelines offer the strongest overall privacy--utility--scalability trade-off in the evaluated setting.
\end{itemize}

\section{Related Works}\label{sec:related_work} 
PCA and population stratification: PCA is a popular technique used to analyze and identify the patterns and relationships within the data. One of the main approaches to employ PCA is to detect population substructures in genetic data. Lee et al.\cite{Lee2009PCA} initially apply PCA to the genotype data of a population, select the most significant principal components by using Tracy-Widom distribution, and finally use generic clustering algorithms to determine the population substructure.

\subsection{Hierarchical Clustering (HC) in Population Stratification:}
In several studies on population stratification, Principal Component Analysis (PCA) was typically applied first to reduce the dimensionality of genetic data, making it easier to identify patterns and reduce noise. Traditionally, following PCA, K-Means clustering were employed to identify population subgroups. However, Hierarchical Clustering (HC) was used in some studies due to its ability to uncover complex, nested population structures without requiring the number of clusters to be predefined. 
Hierarchical Clustering is favored over K-Means for population stratification for several reasons:
\begin{itemize}
    \item No Predefined Number of Clusters: Unlike K-Means, HC does not require the number of clusters to be predefined, making it more flexible for datasets with unknown cluster counts.
    \item Handling Hierarchical Structures: Population data often exhibits nested or hierarchical structures, which HC can reveal through its dendrogram representation.
    \item Uneven Cluster Sizes: HC is more adept at handling clusters of different sizes and shapes, whereas K-Means assumes spherical clusters of equal size.
    \item Interpretability: HC provides a dendrogram, offering a clear visual representation of relationships between populations, which is valuable for understanding the data's underlying structure.
    \item Flexibility with Data: HC can use various distance measures, making it versatile for different types of population data, including genetic datasets.
\end{itemize}

Bouaziz et al. \cite{Bouaziz2012SHIPS} applied PCA to simplify data before using Spectral Hierarchical Clustering (SHIPS) to infer population structure. The Spectral Hierarchical Clustering (SHIPS) used in this study was applied to reduce population structure complexity. Hierarchical clustering was chosen for its ability to handle data with nested structures, which is common in genetic studies. 
Jafarzadegan et al.\cite{Jafarzadegan2019Combining} combined PCA with Hierarchical Clustering to reveal population substructure in a dataset with unclear groupings. Hierarchical clustering was used here after PCA to explore population substructure in an unbiased way. The advantage of hierarchical clustering in this study is that it provides a dendrogram, which helps researchers visualize the relationships between subgroups in the population, which is crucial when there is uncertainty about the number of clusters.
Maugeri et al. \cite{Maugeri2021HCPC} analyses regional COVID-19 dynamics in Italy with a Hierarchical Clustering on Principal Components (HCPC) workflow. They first reduced 36 highly correlated epidemiological indicators to three principal components, then applied Ward’s agglomerative hierarchical clustering on the PC scores and finally consolidated the partition with k-means to stabilize membership assignments. This PCA → HC (+ K-means) pipeline was chosen to eliminate multicollinearity,exploit Ward’s variance-based linkage for an interpretable dendrogram that suggests the optimal number of clusters, and to refine the solution for greater within-cluster homogeneity—altogether yielding a robust, interpretable grouping of regions with similar epidemic trajectories.
Tonkin-Hill et al.\cite{Tonkin-Hill2019Fast} used Hierarchical Bayesian Clustering, leveraging hierarchical clustering for better interpretation of genetic data. This study applied Hierarchical Bayesian Clustering to interpret population structure in genetic data. The hierarchical approach was beneficial because it could model more complex structures in the data, allowing for better inference of subpopulations compared to traditional K-Means.

In conclusion, combining PCA with hierarchical clustering techniques has proven to be an effective approach for analyzing population stratification, as it reduces data complexity and reveals intricate population structures without the limitations of predefined cluster counts.

\subsection{Gaussian Mixture Models (GMM) in Population Stratification:}
Several studies have utilized Gaussian Mixture Models (GMM) for population stratification due to their ability to probabilistically model the underlying genetic structure of populations. These models are particularly useful when dealing with complex genetic datasets where population subgroups may exhibit multi-modal distributions.

Budiarto et al. \cite{Budiarto2021Gaussian} proposed a GMM-based approach to estimate population stratification from Single Nucleotide Polymorphism (SNP) data, combined with Principal Component Analysis (PCA) for dimensionality reduction. The GMM was used to model the genetic structure probabilistically, capturing complex variations that traditional clustering methods might overlook. The authors chose GMM for its flexibility in identifying subgroups with varying genetic distributions, which is crucial for accurate population stratification in genomic studies.    
Gardner et al. \cite{Gardner2022DA-GMM} applied a domain-adapted GMM approach to population-based structural health monitoring. Although primarily focused on structural health, this study demonstrated the flexibility of GMM in handling different population structures by adapting the model learned from one population to another. The study highlighted GMM's ability to model varying distributions in population data, making it well-suited for complex population-level analyses, even in non-genomic contexts.

The use of GMM in these studies is motivated by its ability to:
\begin{itemize}
    \item \textbf{Model Complex Population Structures}: GMM captures multi-modal genetic distributions, enabling more accurate population stratification, especially in datasets with overlapping or hierarchical population structures.
    \item \textbf{Probabilistic Clustering}: Unlike traditional clustering methods, GMM allows for probabilistic assignments, making it possible to identify subgroups with varying degrees of membership, which is particularly useful when the number of populations is unknown.
    \item \textbf{Flexibility}: GMM can easily adapt to varying population structures, making it a versatile tool for both genomic and non-genomic population-level analyses.
\end{itemize}

\subsection{Density-Based Spatial Clustering of Applications with Noise (DBSCAN)}
Density-Based Spatial Clustering of Applications with Noise (DBSCAN) is a density-based clustering algorithm that groups points into clusters when they are located in sufficiently dense regions, while labeling isolated points as noise (outliers). Unlike K-means, DBSCAN does not require the number of clusters $k$ to be specified in advance, and unlike many hierarchical clustering (HC) approaches, it does not rely on a single global tree cut to define cluster membership. These properties make DBSCAN particularly attractive for genetic population structure analyses, where clusters may be irregularly shaped, unequally sized, and contaminated by outliers.

In the population stratification literature, DBSCAN is often applied after dimensionality reduction (e.g., PCA or PCA$\rightarrow$UMAP), where genetic substructure becomes visible as compact groups separated by sparse regions.

Note that none of the aforementioned studies in this section takes in consideration the privacy of the dataset participants (especially, in collaborative studies), which is one of the main goals of our work. Also, all of them utilized PCA in their studies while our framework considers applying clustering algorithms without dimensionality reduction techniques alongside with the PCA utilized approaches.

Our contribution is a privacy-preserving selection layer for choosing the clustering workflow of the population stratification and not introducing a new DP/MPC population-stratification algorithm. Prior DP works like Dervishi et.al \cite{Dervishi2023Privacy}, typically fix the pipeline (e.g., PCA + K-Means with preset $K$) and then execute a secure protocol; By contrast, our framework lets the server recommend the algorithm-hyperparameter pair from privacy-protected shares, so collaborators are not forced to rely on a one-size-fits-all configuration. Our preliminary work \cite{Ghasemian2025PrivacyCollaborative} introduced the core idea of privacy-preserving algorithm and hyperparameter selection for collaborative population stratification under a public-PCA setting. The present paper generalizes that selection-layer idea to a distributed-PCA setting and broadens both the candidate algorithm space and the evaluation.

\section{Background}\label{sec:background}
In this section we review some background and definitions of different clustering algorithms and clustering evaluation metrics as well as the local differential privacy. 
\subsection{Clustering algorithms for population stratification} \label{clustering_algorithms}

This study examines clustering algorithms commonly used in population stratification. 

K-Means is one of the simplest and most widely used unsupervised learning algorithms for clustering and population stratification. It partitions the dataset into $K$ distinct, non-overlapping clusters by minimizing the sum of distances between the data points and their respective cluster centroid, which serves as the representative of the cluster~\cite{wu2021secure,mohassel2020practical}. 
Gaussian Mixture Models (GMM) accommodates clusters with different sizes and correlations by assuming that the data points are generated from a mixture of several Gaussian distributions. Each cluster corresponds to a different Gaussian distribution, making this algorithm highly flexible when dealing with complex cluster structures~\cite{hamidi2019privacy,lin2005privacy}. 
Hierarchical clustering is another algorithm used in some studies for population stratification which builds a tree of clusters and normally does not require a pre-specified number of clusters. Hierarchical clustering can be either agglomerative (bottom-up approach) or divisive (top-down approach). It is particularly useful for hierarchical data and can provide insights into the data structure at different levels of granularity. However, it is computationally expensive for large datasets and is not scalable. The results can also vary significantly based on the choice of the linkage criterion, which determines how the distance between clusters is measured~\cite{meng2019private,xu1998distribution,everitt2011cluster,ester1996density,fukunaga1975estimation}. 

\subsection{Evaluation Metrics for Clustering Algorithms}
Here we outline the evaluation metrics used later in section \ref{sec:evaluation} to assess clustering quality. To measure the performance of our approach, we use the following metrics, each selected for its capability to capture various dimensions of clustering quality and privacy preservation:

\textbf{Silhouette Score:} This metric evaluates how well each object lies within its cluster, with a focus on cohesion and separation. This score ranges from -1 to 1, where a high value indicates that the object is well matched to its own cluster and poorly matched to neighboring clusters. \textit{Higher} Silhouette Scores signify better-defined clusters, making it a valuable metric for assessing the distinctiveness of the clusters formed by the algorithm.

\textbf{Calinski-Harabasz Index (CH):} The Calinski-Harabasz Index is a method for evaluating the quality of a clustering. It measures the ratio of the sum of between-clusters dispersion and of within-cluster dispersion for all clusters. The \textit{higher} the value of the CH index, the better the clustering model, as it indicates a higher degree of separation between clusters.

\textbf{Davies-Bouldin Index (DB):} The Davies-Bouldin Index is an internal evaluation scheme for clustering algorithms where a \textit{lower} DB index relates to a model with better separation between the clusters. This index signifies the average 'similarity' between clusters, where similarity is a measure that compares the distance between clusters with the size of the clusters themselves.

\subsection{Principal Component Analysis (PCA)}

The objective of Principal Component Analysis (PCA) \cite{Abdi2010Principal, Pedregosa2011Scikit} is to identify the main sources of variation within high-dimensional, inter-correlated datasets by generating new orthogonal variables called principal components. These principal components are linear combinations of the original variables. PCA achieves this by computing the eigenvectors and eigenvalues from the covariance matrix, then sorting the eigenvectors based on the amount of explained variance, in descending order of their corresponding eigenvalues. Typically, only the top eigenvectors are utilized to transform the data, and these are referred to as the principal components. The number of principal components derived from PCA can be adjusted.
\subsection{Local Differential Privacy}

Local Differential Privacy (LDP)~\cite{costello2015geppetto,davidson2011on, Duchi2013Local, Kairouz2014Extremal} is a more restricted form of traditional differential privacy~\cite{Dwork2008Differential, davidson2011provenance}. Unlike traditional differential privacy, LDP does not rely on a trusted third party and provides a higher level of data protection for users. In LDP, each user modifies their own data before sharing them with a data aggregator. The aggregator only sees the perturbed data, ensuring privacy. 
An algorithm \(A\) satisfies $\epsilon$-local differential privacy ($\epsilon$-LDP) if, for any input values \(v1\) and \(v2\): 
\begin{math}
  Pr[A(v1)=y]\leq e^\epsilon Pr[A(v2)=y]  
\end{math},
This condition holds true for all possible outputs of the algorithm \(A\).

The main challenge is to determine the amount of noise to add to achieve LDP, while still maintaining a good level of utility. Several different mechanisms have been developed in the DP field to solve this problem, the most well-known of which is the Laplacian mechanism \cite{Dwork2006Calibrating}. For any numerical function $f(x) : \mathbb{R} \to \mathbb{R}$, $F(X)$ satisfies $\epsilon$-LDP if we add Laplacian noise as follows:
\begin{math}
F(x) = f(x) + \operatorname{Lap}\Big(\frac{s}{\epsilon}\Big),
\end{math}
where $s$ is the sensitivity of the function $f$. Sensitivity captures the maximum amount of change that a single data point can cause in the worst case to the output of the function $f$. In our work, we use the $l_1$ sensitivity to add Laplacian noise \cite{Dwork2014TheAlgorithmic}. By definition,
\begin{math}
s = \max_{x,x'\in\mathbb{R}} \| f(x) - f(x') \|,
\end{math} 
and $\operatorname{Lap}(\lambda)$, where $\lambda=\frac{s}{\epsilon}$, denotes sampling from a Laplace distribution with scale $k$ and with a probability density function:
\begin{math}
\textit{pdf}(x \mid \mu, \lambda) = \frac{1}{2\lambda} \exp\Bigl(-\frac{\lvert x - \mu\rvert}{\lambda}\Bigr),
\end{math}
where $\mu$ is a location parameter (typically $\mu$ = 0 for a symmetric distribution).
\section{System and Threat Models}\label{sec:system_model}

\begin{table*}[h]
\centering
\caption{Table of symbols and notations. }
\label{tab:symbols_notations}
\begin{tabular}{|l|l|}
 \hline
Symbol & Description  \\
\hline
S & Server\\
$D_i$ & Dataset of each party $i$\\
$ND_i$ & Noisy data of each party $i$ produced as a result of Laplace Noise\\
$S_i$ & LDP\_processed data of party $i$\\
$Q^{\mathrm{dist}}$ & Global PCA basis learned by the server from aggregated local PC summaries.\\
$\epsilon$ & epsilon, Privacy Parameter \\
$V_i$ & Local top-$r$ PCA basis at party $i$\\
K & Number of clusters \\
CH & Calinski-Harabasz Index \\
DB & Davies-Bouldin Index\\
$n^i$ & Number of individuals in $D_i$ \\
\hline
\end{tabular}    
\end{table*}

\subsection{System Model}
Our collaborative population stratification framework consists of two main entities: an honest-but-curious server and at least two parties (researchers). The honest-but-curious server assists the parties in determining the best combination of clustering algorithm and parameters for their collaborative population stratification task. If the parties decide to use PCA, each party computes a local top-$r$ PCA basis and send only this low-rank summary to the server. The server aggregates these summaries to construct a shared distributed-PCA basis $Q^{\mathrm{dist}}$, which is then broadcast to all parties. Parties then execute one of three local pipelines, each providing an $\epsilon$-LDP guarantee.  Finally, the server aggregates the resulting shares to perform clustering and parameter selection.  Table~\ref{tab:symbols_notations} lists the symbols and notations used throughout the paper. 
Formally, the components are:
\begin{itemize}
  \item \textbf{Parties (Researchers):} A collection of $n$ institutions, each holding a private dataset $D_i \subset \mathbb{R}^d$ of high-dimensional genetic or phenotypic measurements. Each party applies one of the three LDP scenarios to produce an LDP‐processed data $S_i$ and send that to the server which helps the server identify a suitable clustering algorithm–hyperparameter configuration. Their immediate goal in this framework is to obtain a privacy-preserving recommendation of a strong clustering algorithm-hyperparameter pair for collaborative population stratification, while keeping raw genotypes private.
  \item \textbf{Honest-but-curious Server:} The server aggregates transformed or noisy data from all parties, performs clustering, and then returns the recommended algorithm along with the optimal parameter choices to the parties.
  \item \textbf{Optional Distributed-PCA Model:} 
  When a pipeline requires PCA, the parties must share a common projection. Each party standardizes its features and computes a local top-$r$ basis $V_i$. (or projector \(P_i = V_i V_i^\top)\)). Parties send only \(V_i / P_i\) (optionally weighted by $n_i$) to the coordinator, which aggregates the projectors (or uses the small Gram matrix of concatenated bases) to extract the global top-$r$ eigenvectors $Q^{\mathrm{dist}}$. The server broadcasts $Q^{\mathrm{dist}}$ to all parties. No raw genotypes are exchanged.  
\end{itemize}

We evaluate the empirical $\epsilon$-LDP privacy behavior of all three pipelines in Section \ref{sec:evaluation}, while the formal privacy guarantee is analyzed in section \ref{sec:privacy}.

\subsection{Threat Model}
In our setting, we assume that each research party is honest: they hold legitimate genomic or phenotypic datasets and follow the protocol as specified. The only adversary is the server, which we model as "\textit{honest-but-curious}": it correctly performs all computations—aggregating local PC summaries to obtain $Q^{\mathrm{dist}}$, collecting each party’s differentially private share, running the clustering and validation procedures, and returning its recommendation, but it may inspect the metadata (i.e. the noisy feature or PCA‐score vectors) in an attempt to infer information about individual participants.

Genomic datasets are known to be susceptible to several types of inference attacks. In a membership-inference attack \cite{Homer2008Resolving, Wang2009Learning}, an adversary who has access to an individual’s genomic profile tries to determine whether that person’s data were included in a particular dataset. In an attribute-inference attack \cite{Humbert2013Addressing}, the adversary leverages auxiliary information to predict sensitive traits of participants (such as disease status or ancestry). In de-anonymization attacks \cite{gymrek2013identifying}, the goal is to link an identity—often available from public records—to a record in the dataset. Of these, membership inference is the most relevant in our considered setting: our \textit{honest-but-curious} server might try to use the noisy shares $S_i$ to decide whether a particular individual contributed data to one of the parties.

To combat these attacks, we require that each party’s transmitted share satisfy $\epsilon$-local differential privacy. By calibrating Laplace noise to the global sensitivity of each coordinate—whether in the raw $d$-dimensional space or in the $r$-dimensional PCA projection—we ensure that observing $S_i$ reveals almost no information about the presence or absence of any single record beyond what is permitted by $\epsilon$. 

Our model relies on three core assumptions. First, each party correctly implements its LDP pipeline—using the distributed-PCA model $Q^{\mathrm{dist}}$ when required, computing sensitivities from known feature ranges, and sampling noise from the Laplace distribution with the agreed parameters. Second, there is no collusion between any party and the server; parties never share raw data with the server. Third, feature ranges (and hence sensitivities) for both raw and PCA-transformed data can be safely bounded, enabling correct noise calibration. Given these conditions, any further clustering, metric computation, or algorithm‐and‐parameter selection performed by the server is on LDP-protected data and incurs no additional privacy loss.

In the following, we list the specific adversary goals for each of the three settings we present:
\paragraph{PCA\,$\rightarrow$\,Noise}  
The server observes only $r$ noisy components.  Although dimensionality reduction limits exposure, a semi‐honest server may still attempt inference using the released noisy low-dimensional scores together with auxiliary information.

\paragraph{Noise\,$\rightarrow$\,PCA}  
Noise is first injected in the $d$‐dimensional space; subsequently projecting onto top PCs concentrates signal‐to‐noise ratio along principal axes.  An adversary could thus focus inference on retained dimensions where noise impact is minimal.

\paragraph{Noise-Only}  
Full $d$‐dimensional noisy data provides maximal visibility.  In this setting, the server can apply reconstruction or de‐noising techniques on each feature independently, increasing membership‐inference risk when $\epsilon$ is large enough.

\section{Proposed Framework}\label{sec:proposed_work}
\begin{figure*}[t]
   \centering
   \includegraphics[width=0.9\linewidth]{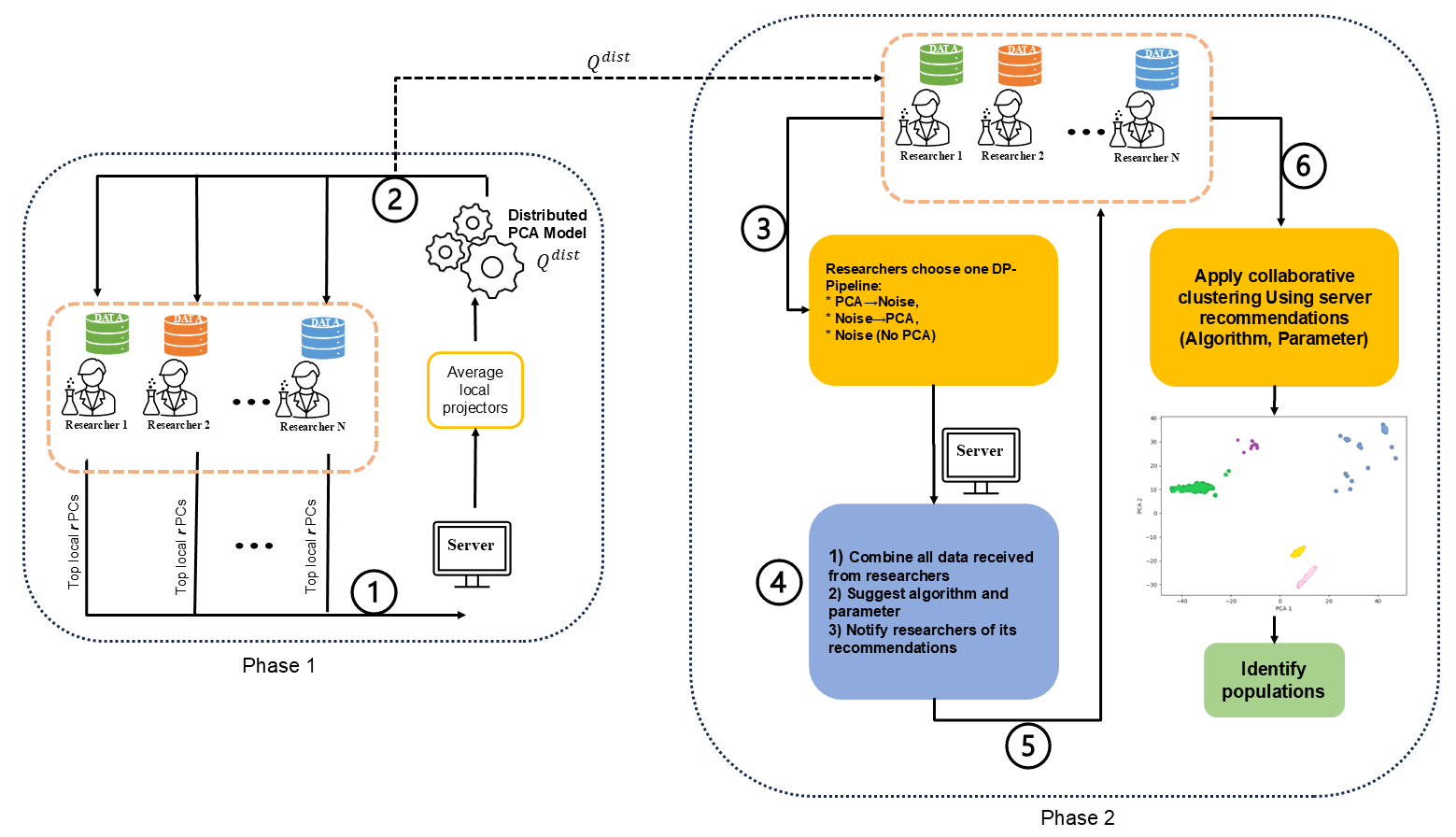}
   \caption{System model for privacy-preserving collaborative population stratification with distributed-PCA. 
Phase~1: each researcher keeps raw data local, computes top-$r$ local PCs, and sends only the resulting low-rank projectors to the honest-but-curious server (step 1). The server averages these projectors to obtain the distributed-PCA basis $Q^{\mathrm{dist}}$ and broadcasts $Q^{\mathrm{dist}}$ back to all parties (step 2). 
Phase~2: researchers choose one of three $\epsilon$-LDP pipelines (PCA\,$\rightarrow$\,Noise, Noise\,$\rightarrow$\,PCA, or Noise-Only) to generate sanitized data shares and transmit them to the server (step 3). The server concatenates all received shares, evaluates multiple clustering algorithms and hyperparameter settings using internal validation metrics (step 4), and recommends the best (algorithm-hyperparameter) configuration (step 5). the Downstream stratification analysis is then carried out by the researchers (parties) using that recommended configuration, without ever revealing raw genotypes to the server (step 6).}
  \label{fig:system_model}
\end{figure*}

Our framework implements and compares several privacy‐preserving pipelines in a collaborative clustering setting. Because of the available resources and (privacy and utility) preferences across parties, we introduce three complementary pipelines: two that combine PCA with Laplace noise in different orders, and a third that omits dimensionality reduction entirely. All three pipelines satisfy the $\epsilon$-LDP guarantee, but each occupies a different point in the privacy $\leftrightarrow$ utility $\leftrightarrow$ complexity triangle. 
Figure \ref{fig:system_model} outlines the overall workflow and the following subsections provide the details.

\subsection{Distributed-PCA Model} \label{sec:DPCA}
When a pipeline calls for PCA, all parties must project into the same coordinate system. Following a distributed-PCA approach \cite{Fan2019DistributedPCA}, each party computes a locally centered top-$r$ basis $V_i$ from its own dataset. The server concatenates these local bases into $Y = [V_1,...V_m]$, forms the Gram matrix $Gram = Y^T Y/m$, and extracts the $Q^\mathrm{dist}$ from the top eigenvectors of $Gram$, mapped back to features space. This basis-construction step is a coordination phase used by the PCA-based release pipelines; because of PCA transformation, no raw data are exchanged in this process.

\subsection{Pipeline 1: PCA $\rightarrow$ Noise}

In the first pipeline, each party projects its standardized data $\tilde D_i\in\mathbb{R}^{n_i\times d}$ onto the top $r$ components of the distributed-PCA basis $Q^\mathrm{dist}$, producing a score matrix $P_i\in\mathbb{R}^{n_i\times r}$. Next, it perturbs each coordinate $P_{i,k,j}$ by adding independently drawn Laplace noise with scale $\Delta_j/\epsilon$, where $\Delta_j = \max_k P_{i,k,j} - \min_k P_{i,k,j}$ is the sensitivity of the $j$th component. The noisy scores $\widehat P_i$ constitute that party’s private share, which is sent to the server.

In practice, adding noise after projection into a low-dimensional subspace typically yields higher utility than perturbing the original high-dimensional data. This is consistent with standard differential privacy theory: the Laplace mechanism adds noise proportional to the $\ell_1$-sensitivity of each coordinate and a reduced dimension $r\ll d$ generally implies smaller per-coordinate sensitivity and therefore less injected noise overall \cite{Dwork2014The}. Chaudhuri et al.\ \cite{chaudhuri2013a} empirically demonstrate that their “PPCA” output-perturbation method for private PCA achieves substantially better subspace estimation than raw-data mechanisms, particularly when the ambient dimension $d$ is large.

From a clustering standpoint, low-dimensional, noisy PC‐scores preserve the dominant axes of variation that encode subpopulation structure. Recent work by He et al. \cite{He2025Differentially} formally shows that a private PCA step can circumvent the curse of dimensionality, yielding synthetic low-dimensional representations whose Wasserstein-distance utility degrades only with the intrinsic subspace dimension $r$ rather than $d$. Likewise, Liu et al. \cite{liu2022dppca} prove near-optimal error bounds for private PCA under minimal sample requirements, reinforcing that concentrating the privacy budget into the top $r$ components preserves more signal‐to‐noise than distributing it across all $d$ raw features.

\subsection{Pipeline 2: Noise $\rightarrow$ PCA}

The second pipeline reverses the order of transformation and noise. Each party first adds Laplace noise to every raw SNP coordinate in $\tilde D_i$, thereby satisfying $\epsilon$-LDP immediately . It then uses the same distributed-PCA model $Q^\mathrm{dist}$ described above (in Section \ref{sec:DPCA}), to obtain a reduced matrix $P_i^{\mathrm{noisy}}\in\mathbb{R}^{n_i\times r}$. As in the first pipeline, these noisy principal‐component scores serve as the local LDP share that is transmitted to the server for algorithm-hyperparameter selection phase.

\subsection{Pipeline 3: Noise‐Only (No PCA)}

In the simplest pipeline, parties inject Laplace noise independently into every released raw feature of $\tilde D_i$ and share the full $n_i\times d$ noisy matrix with the server. No dimensionality reduction is performed, and the server clusters in the raw feature space. By omitting PCA, this pipeline avoids projection overhead, but it requires clustering in the original high-dimensional space, which increases communication cost and server-side runtime (Section \ref{sec:complexity}). In our experiments, it also yields weaker clustering utility and higher MIA power than the PCA-based pipelines.

\subsection{Server‐Side Clustering and Parameter Tuning}\label{sec:scoring}

After receiving each party’s LDP-protected output, the server concatenates all shares into an aggregate matrix $S$ of size $N\times d'$ (where $d'=r$ for the first two pipelines and $d'=d$ for Noise-Only). The server then evaluates four candidate clustering algorithms--K-Means, Gaussian Mixture Models, DBSCAN and Agglomerative Clustering--over a range of algorithm-specific hyperparameter settings. For DBSCAN, the server searches over density-specific hyperparameters (Eps and minPoints) rather than a target number of clusters. For each (algorithm, parameter) combination, the server computes three internal validation metrics (silhouette score, Calinski–Harabasz index, Davies–Bouldin index), rescales them to $[0,1]$, and combines them into a composite quality score
\begin{equation}
\label{eq.score}
  \mathrm{Score} = W_{sil}\,\tilde{s}_{\mathrm{sil}} + W_{CH}\,\tilde{s}_{\mathrm{CH}} + W_{DB}\,(1 - \tilde{s}_{\mathrm{DB}}).
\end{equation}
Weight values in Eq.\ref{eq.score} are data and experiment dependent and can be adjusted accordingly. In our experiments, silhouette is given the largest weight ($W_{sil} = 0.5$) because it simultaneously measures cohesion and separation, whereas the CH and DB indices capture complementary notions of cluster scatter, and thus each receives weight $0.25$. The (algorithm, parameter) setting with the highest composite score is returned to all parties as the recommended clustering method and corresponding hyperparameter. Since all operations beyond local noise injection are pure post-processing of LDP-protected shares, they incur no additional privacy loss.

By combining these three complementary pipelines with an automated, privacy-preserving selection procedure, the framework allows collaborators to tune the privacy–utility–complexity trade-off to their constraints while still converging on a near-optimal clustering solution. For PCA-based pipelines, communication scales as $O(n \cdot r)$ ($n$ samples, $r$ PCs) instead of $O(n \cdot d)$ ($d$ SNPs), thereby reducing both network cost and server runtime relative to clustering directly in the full $d$-dimensional space.
\section{Privacy Analysis}\label{sec:privacy}
Our framework enforces LDP via the Laplace mechanism, which perturbs every released statistic by adding noise drawn from a Laplace distribution whose scale is proportional to that statistic’s $\ell_1$–sensitivity.

In the \emph{PCA\,$\rightarrow$\,Noise} pipeline, each party first projects its standardized feature matrix $\mathbf D_i\in\mathbb R^{n_i\times d}$ onto the top $r$ components of the distributed-PCA model $Q^\mathrm{dist}\in\mathbb R^{d\times r}$, yielding score matrix $\mathbf P_i = \mathbf D_i\,Q^{\mathrm{dist}}$.  The $\ell_1$–sensitivity of the $j$th principal‐component coordinate is
\begin{equation}
\label{eq.Sensitivity_j}
    \Delta_j \;=\;\max_n P_{i,n,j} \;-\; \min_n P_{i,n,j}
\end{equation}
 
and the party releases independently perturbed entries 
\begin{equation}
\label{eq.ReleasedNoisy}
  \widehat P_{i,n,j}
    = P_{i,n,j} \;+\; \mathrm{Lap}\bigl(\Delta_j / \epsilon_j\bigr),
    \qquad
    \sum_{j=1}^r \epsilon_j = \epsilon  
\end{equation}

which, by the sequential composition theorem, guarantees $\epsilon$–LDP. Since all subsequent clustering and metric computations operate solely on these noisy scores, they incur no additional privacy loss.

In the \emph{Noise\,$\rightarrow$\,PCA} pipeline, noise is instead added directly to each raw feature $D_{i,n,p}$ according to its feature-range sensitivity
\begin{equation}
\label{eq.delta_p}
  \Delta_p = \max_nD_{i,n,p} \;-\; \min_nD_{i,n,p}  
\end{equation}

by publishing
\begin{equation}
   \widetilde D_{i,n,p}
    = D_{i,n,p} \;+\; \mathrm{Lap}\bigl(\Delta_p / \epsilon_p\bigr),
    \qquad
    \sum_{p=1}^d \epsilon_p = \epsilon 
\end{equation}

This provides $\epsilon$–LDP for the full $d$-dimensional dataset. Any subsequent deterministic projection $\widetilde P_i = \widetilde D_i\,\mathbf Q^\mathrm{dist}$ preserves this privacy guarantee as a post-processing step.

In the \emph{Noise-Only} pipeline, parties perturb each raw feature exactly as above but skip PCA, so the released matrix $\widehat{\mathbf D}_i$ of size $n_i\times d$ immediately satisfies $\epsilon$–LDP under the Laplace mechanism.

Because the PCA\,$\rightarrow$\,Noise pipeline reduces the release dimension from $d$ to $r\ll d$, it concentrates the same privacy budget into fewer coordinates, which empirically leads to higher utility for a fixed $\epsilon$. Across all three pipelines, the total privacy loss per record is bounded by $\epsilon$, and the server’s subsequent automated algorithm and parameter selection is pure post-processing, incurring no extra privacy cost.
 
Our framework also conducts a power analysis of membership inference attacks, using Euclidean distance to quantify the potential risk posed by an adversarial server. We evaluate the membership inference risk for a victim $v$ belonging to a party’s dataset $D_i$, assuming that the released metadata (produced by any of the three pipelines) has $d$ dimensions. We first construct a control group $G$ containing $|G|$ individuals who are guaranteed not to be in $D_i$, sampled at random from the pool of all individuals outside $D_i$. For each target individual $t\in G$, we compute the Euclidean distance between the metadata of $t$ and the metadata of every individual in $D_i$ over the $d$ dimensions, and record the minimum distance for $t$. We then define a Euclidean-distance threshold $T$ that corresponds to a 5\% false positive rate, i.e., 95\% of the individuals in the control group $G$ are correctly identified as not belonging to $D_i$. This threshold serves as the decision boundary beyond which an individual is deemed unlikely to be a member of the researcher’s dataset.

\section{Evaluation}\label{sec:evaluation}
We evaluate the proposed framework on two real genomic datasets: openSNP (942 individuals, 28,396 SNPs) \cite{Greshake2014openSNP, openSNP2025News} and 1000 Genomes Phase III (600 individuals, approximately 37,000 SNPs) \cite{1000genom}. For openSNP, we preprocess the genotype matrix with PLINK by removing SNPs with minor-allele frequency below 0.01 and SNPs with missingness above 5\%. We repeat the same evaluation on 1000 Genomes Phase III to verify that the observed trends are not dataset-specific. When PCA is used, the parties construct a shared low-dimensional representation through the distributed-PCA procedure described earlier. Unless otherwise noted, we use $\epsilon$-LDP, retain $r=2$ principal components in the PCA-based pipelines, and average each point over 20 runs. For a fair comparison, the public-data PCA baseline and the distributed-PCA experiments use the same collaborative cohort size: 710 samples for openSNP and 450 samples for 1000 Genomes. The difference between the two settings is therefore only how the PCA basis is constructed: from a public dataset in the baseline, and from party-local summaries in the distributed-PCA setting. The public-data PCA baseline corresponds to the setting studied in our preliminary work in \cite{Ghasemian2025PrivacyCollaborative}, where the server used a public dataset to provide a common PCA basis. 

In addition to evaluating the proposed distributed-PCA framework, we also compare against a baseline setting in which the server constructs the PCA model using party shares together with a public dataset. Accordingly, Figure \ref{fig:DPCA_metrics_epsilon} reports the privacy--utility results for the distributed-PCA framework, with subfigures (a)--(c) for openSNP and (d)--(f) for 1000 Genomes. Figure \ref{fig:metrics_epsilon} reports the corresponding results for the public-data PCA baseline.

\subsection{Server Recommendation and Collaborative Clustering Quality}
Table~\ref{tab:server_suggestion} summarizes the clustering configurations (algorithm-hyperparameter pairs) recommended by the server for both the openSNP and 1000 Genomes datasets under a privacy budget of $\epsilon = 3.0$. For each LDP pipeline, the server aggregates the noisy shares and evaluates K-Means, Gaussian Mixture Models (GMM), and Agglomerative Clustering (HC) and DBSCAN over algorithm-specific hyperparameter settings (cluster counts $K$ for K-Means, HC and GMM and, \textit{Eps} and \textit{minPoints} for DBSCAN). For every candidate configuration, the server computes the weighted composite of silhouette, CH, and DB indices described in Section~\ref{sec:scoring}, and reports the configuration with the highest overall selection score. 

Table \ref{tab:combined_clustering} shows the clustering quality obtained when these recommended configurations are applied to the combined noisy data. The correct interpretation is that the server-side recommendation layer consistently identifies one of the top performing algorithm--hyperparameter settings, rather than a uniquely dominant choice in every case. On openSNP, under both PCA-based pipelines, GMM, HC and K-Means at $K=3$ achieve identical or nearly identical selection scores in table \ref{tab:server_suggestion} and the same downstream performance score in table \ref{tab:combined_clustering}. On 1000 Genomes, the same pattern holds: the strongest configurations under the PCA-based pipelines are concentrated at $K=3$, with K-Means and HC achieving the top downstream performance and GMM remaining close behind. By contrast, Noise-Only yields substantially weaker clustering quality on both datasets, with lower silhouette and CH values and higher DB values.  

DBSCAN provides an informative contrast to the centroid- and linkage-based methods. On the two PCA-based pipelines, it usually returns a valid clustering, but its quality is consistently below that of K-Means and HC. On openSNP, DBSCAN achieves performance scores of 0.60 under Noise$\rightarrow$PCA and 0.63 under PCA$\rightarrow$Noise, compared with 0.71 for GMM, HC, and K-Means. On 1000 Genomes, DBSCAN reaches 0.67 under both PCA-based pipelines, whereas HC and K-Means reach 0.71. This suggests that density-based clustering can recover meaningful structure once the data have been projected into a low-dimensional representation, but in the present datasets it does not match the best centroid- or hierarchy-based solutions. Under Noise-Only, DBSCAN is even less stable and fails to return a usable partition on both datasets, indicating that density estimation in the full noisy feature space is particularly brittle.

Taken together, the two tables underscore two main points. First, despite the differing noise‐injection orders and dimensionalities, the server‐side recommendation layer consistently identifies one of the top performing algorithm-parameter settings.  Second, by exploring three complementary LDP-pipelines, the framework allows collaborators to adopt a strong configuration for their chosen privacy setting without manual parameter tuning. These results support the main motivation of the paper where a commonly used fixed workflow (e.g., PCA + K-means with preset $K$) is not always uniquely optimal, and collaborators benefit from a privacy-preserving selection layer that adapts the clustering configuration to the dataset. 

\begin{table*}[tb]
    \caption{Server Suggestions for Clustering Input Parameters: Recommendations for various clustering algorithms on different pipelines ($\epsilon$ = 3.0). For each LDP pipeline the server reports the algorithm and hyperparameter that maximize a weighted composite of silhouette, CH and DB indices, together with the individual metric values and the resulting selection score. }
    \centering
    \begin{tabular}{|c|l|l|c|c|l|c|c|c|c|}
    \hline
         \textbf{Dataset}&\textbf{Pipeline} & \textbf{Algorithm} & \textbf{$\epsilon$} & \textbf{Hyperparameter} & \textbf{Silhouette} &\textbf{CH} &\textbf{DB} &\textbf{Selection Score} \\
         \hline
          &&GMM & 3.0 & k = 3 & 0.9 & 13020.26& 0.123 & 0.70\\
          openSNP&Noise\,$\rightarrow$\,PCA & HC & 3.0 & k = 3 & 0.90 & 13059.83& 0.12 & 0.70\\
         &&K-Means & 3.0 & k = 3 & 0.9 & 13065.05& 0.12 & 0.70\\
         &&DBSCAN& 3.0 & Eps = 11.06, minPoint = 5 & 0.76 & 4627.74 & 1.52 & 0.56\\
         \hline
          &&GMM & 3.0 & k = 3 & 0.9 & 13025.87 & 0.12 & 0.7 \\
         openSNP&PCA\,$\rightarrow$\,Noise & HC & 3.0 & k = 3 &0.9& 13031.27 &0.12 &0.7\\
         && K-Means  & 3.0 & k = 3 & 0.9 & 12982.1 & 0.12 & 0.7\\
         &&DBSCAN&3.0& Eps = 9.39, minPoint = 5 &0.8& 5773.41& 1.25&0.59\\
         \hline
        &&GMM & 3.0 & k = 5 & 0.07& 12.95& 2.16 &0.17 \\
         openSNP&Noise-Only & HC & 3.0 & k = 2 & 0.42 & 4.9 & 0.44 & 0.23\\
         && K-Means  & 3.0 & k = 2 & 0.08 & 3.35 & 1.16 & 0.11\\
         &&DBSCAN&3.0 & Eps = 340.08, minPoint = 14 &0.2& 8.88& 3.82& 0.36\\
         \hline
         &&GMM & 3.0 & k = 3 & 0.89 & 9631.9& 0.13 & 0.68\\
          1000 Genomes &Noise\,$\rightarrow$\,PCA & HC & 3.0 & k = 3 & 0.89 & 9628.33&  0.131 & 0.68\\
         &&K-Means & 3.0 & k = 3 & 0.89 & 10203.5 & 0.133 & 0.7\\
         &&DBSCAN& 3.0& Eps = 3.31, minPoint = 5 &0.84&7622.7& 0.55& 0.64\\
         \hline
          &&GMM & 3.0 & k = 3 & 0.89 & 9664 & 0.131 &0.68 \\
         1000 Genomes &PCA\,$\rightarrow$\,Noise & HC & 3.0 & k = 3 & 0.89 & 9610.2 &0.131 &0.68\\
         && K-Means  & 3.0 & k = 3 & 0.9 & 10162.2 & 0.133 & 0.69\\
         &&DBSCAN&3.0& Eps = 3.08, minPoint = 5 & 0.84 &7999.05&0.5&0.65\\
         \hline
        &&GMM & 3.0 & k = 5 & 0.0 & 6.31 & 4.84 &0.21 \\
         1000 Genomes &Noise-Only & HC & 3.0 & k = 2 & 0.1 & 1.46 & 0.81 &0.10\\
         && K-Means  & 3.0 & k = 2 & 0.0 & 7.71 & 5.65 &0.25 \\
         &&DBSCAN&3.0& Eps = 365.90, minPoint = 10 & 0.01& 1.74& 2.97&0.16\\
         \hline
    \end{tabular}
    \label{tab:server_suggestion}
   \end{table*}

\begin{table*}[ht!]
    \caption{Results of applying server's suggestion on combined dataset. Silhouette, CH and DB indices measured on the aggregated noisy dataset when collaborators adopt the server’s suggested algorithm and hyperparameter; higher silhouette/CH and lower DB indicate better partitions. }
    \centering
    \begin{tabular}{|c|l|l|c|l|c|c|c|c|}
    \hline
        \textbf{Dataset} & \textbf{Pipeline} & \textbf{Algorithm} & \textbf{Hyperparameter} & \textbf{Silhouette} &\textbf{CH} &\textbf{DB} &\textbf{Performance Score} \\
         \hline
          && GMM & k = 3  & 0.92 & 14365.4 & 0.10 & 0.71\\
         openSNP &Noise\,$\rightarrow$\,PCA & HC & k = 3 & 0.92 & 14365.4 & 0.10 & 0.71\\
         &&K-Means & k = 3 & 0.92 & 14365.4 & 0.10 & 0.71\\
         &&DBSCAN& Eps = 11.06, minPoint = 5  & 0.81 & 5066.43& 1.41& 0.60\\
         \hline
          &&GMM & k = 3 & 0.92 & 14365.4 & 0.1 & 0.71\\
         openSNP &PCA\,$\rightarrow$\,Noise & HC & k = 3 & 0.92 & 14365.4& 0.1 & 0.71\\
         && K-Means & k = 3  & 0.92 & 14365.4 &0.1 & 0.71\\
         &&DBSCAN& Eps = 9.39, minPoint = 5 & 0.84 & 7061.50 & 1.07 & 0.63\\
         \hline
        &&GMM & k = 5 & 0.15 & 73.48  & 3.06  & 0.33 \\
         openSNP &Noise-Only & HC & k = 2 & 0.54 & 7.97 & 0.34 & 0.29\\
         && K-Means & k = 2 & 0.14 & 5.26 & 1.0 &  0.14\\
         &&DBSCAN& Eps = 340.08, minPoint = 14 & - & - & - & -\\
         \hline
          && GMM & k = 3  & 0.92 & 11982.40 & 0.10 & 0.68\\
         1000 Genomes &Noise\,$\rightarrow$\,PCA & HC & k = 3 & 0.92 & 13803.73 & 0.01 & 0.71\\
         &&K-Means & k = 3 & 0.92 & 13803.73  & 0.01 & 0.71\\
        & &DBSCAN& Eps = 3.31, minPoint = 5 & 0.87 & 12067.86 & 0.46 & 0.67\\
         \hline
          &&GMM & k = 3 & 0.92 & 11982.40 & 0.11 & 0.67 \\
         1000 Genomes &PCA\,$\rightarrow$\,Noise & HC & k = 3 & 0.92 & 13803.73 & 0.01 & 0.71\\
         && K-Means & k = 3 &0.92  & 13803.73 & 0.01 & 0.71 \\
         &&DBSCAN& Eps = 3.08, minPoint = 5 & 0.88 & 12123.16 & 0.50 & 0.67\\
         \hline
        &&GMM & k = 5 & 0 & 12.17 & 5.25 & 0.25 \\
        1000 Genomes & Noise-Only & HC & k = 2 & 0.15 & 1.73 & 0.74 & 0.11 \\
         && K-Means & k = 2& 0.0 & 11.76 & 4.59 & 0.24\\
         &&DBSCAN& Eps = 365.90, minPoint = 10 & - & - & - & -\\
         \hline
    \end{tabular}
    \label{tab:combined_clustering}
\end{table*}

\subsection{Privacy–Utility Trade‐off}\label{sec:utility}
\begin{figure*}[!t]
\centering
\setlength{\tabcolsep}{3pt}
\setcounter{panel}{0}

\alt{Six-panel figure showing Silhouette, CH, and DB scores versus $\epsilon$ for openSNP (top row) and 1000 Genomes (bottom row). Distributed-PCA}{%
  \mbox{%
    \begin{tabular}{@{}ccc@{}}

      \refstepcounter{panel}\label{fig:DPCA_Sil_epsilon_OpenSNP}%
      \includegraphics[width=.32\textwidth]{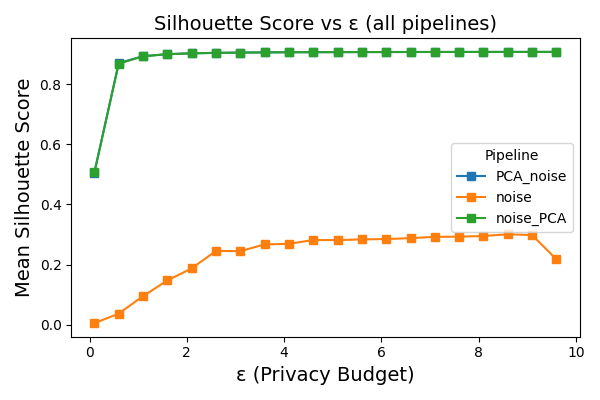} &
      \refstepcounter{panel}\label{fig:DPCA_CH_epsilon_OpenSNP}%
      \includegraphics[width=.32\textwidth]{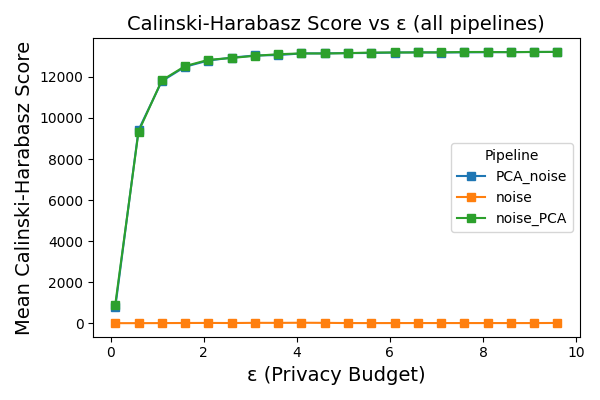} &
      \refstepcounter{panel}\label{fig:DPCA_DB_epsilon_OpenSNP.png}%
      \includegraphics[width=.32\textwidth]{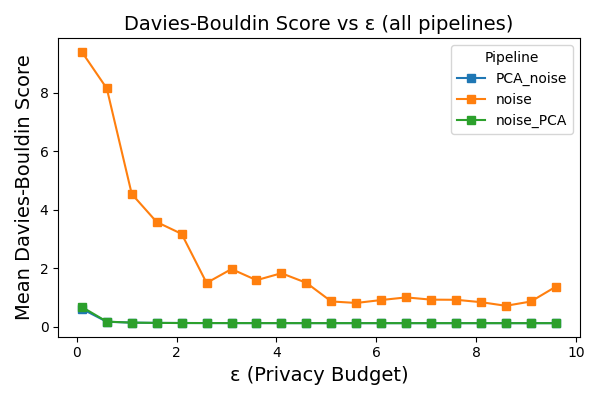} \\
      \small (\ref{fig:DPCA_Sil_epsilon_OpenSNP}) Silhouette vs.\ $\epsilon$ &
      \small (\ref{fig:DPCA_CH_epsilon_OpenSNP}) CH Index vs.\ $\epsilon$ &
      \small (\ref{fig:DPCA_DB_epsilon_OpenSNP.png}) DB Index vs.\ $\epsilon$ \\[0.8ex]

      \refstepcounter{panel}\label{fig:DPCA_Sil_epsilon_1000genome}%
      \includegraphics[width=.32\textwidth]{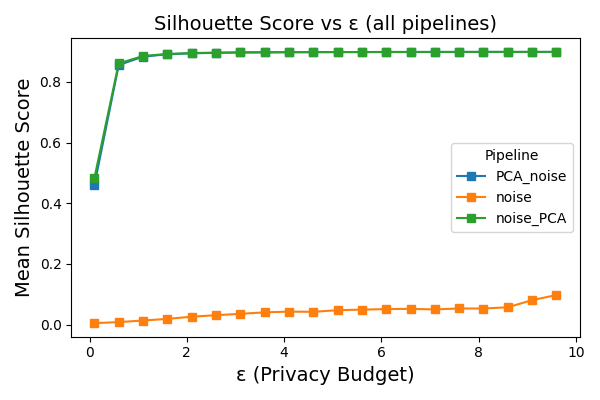} &
      \refstepcounter{panel}\label{fig:DPCA_CH_epsilon_1000genome}%
      \includegraphics[width=.32\textwidth]{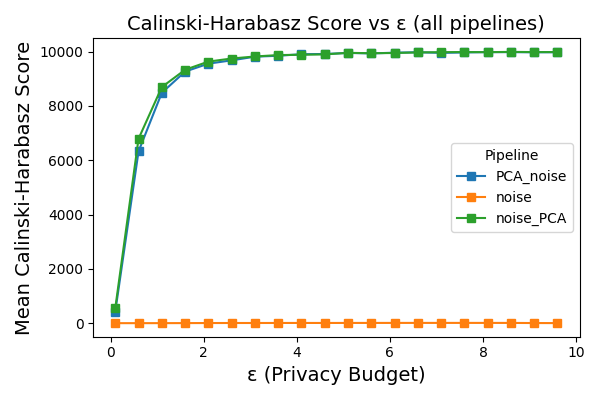} &
      \refstepcounter{panel}\label{fig:DPCA_DB_epsilon_1000genome}%
      \includegraphics[width=.32\textwidth]{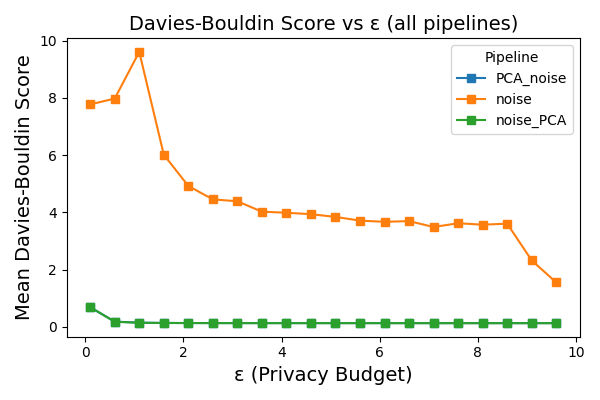} \\
      \small (\ref{fig:DPCA_Sil_epsilon_1000genome}) Silhouette vs.\ $\epsilon$ &
      \small (\ref{fig:DPCA_CH_epsilon_1000genome}) CH Index vs.\ $\epsilon$ &
      \small (\ref{fig:DPCA_DB_epsilon_1000genome}) DB Index vs.\ $\epsilon$ \\

    \end{tabular}%
  }%
}

\caption{Privacy--utility trade-off under the proposed distributed-PCA framework. Subfigures (a)--(c) show openSNP results and subfigures (d)--(f) show 1000 Genomes results.}
\label{fig:DPCA_metrics_epsilon}
\end{figure*}

Figure \ref{fig:DPCA_metrics_epsilon} presents the privacy--utility trade-off for the proposed distributed-PCA framework. As expected, clustering quality improves as the privacy budget $\epsilon$ increases: silhouette and Calinski--Harabasz (CH) generally increase, while Davies--Bouldin (DB) decreases. Across both datasets, the two PCA-based pipelines clearly outperform Noise-Only. Among them, PCA$\rightarrow$Noise generally provides the strongest or near-strongest utility, while Noise$\rightarrow$PCA remains close behind. These results show that the distributed-PCA procedure preserves the utility advantage of low-dimensional collaborative clustering without relying on a public dataset.

\begin{figure*}[!t]
\centering
\setlength{\tabcolsep}{3pt}
\setcounter{panel}{0}

\alt{Six-panel figure showing Silhouette, CH, and DB scores versus $\epsilon$ for openSNP (top row) and 1000 Genomes (bottom row). Centralized PCA}{%
  \mbox{%
    \begin{tabular}{@{}ccc@{}}

      \refstepcounter{panel}\label{fig:Sil_epsilon_OpenSNP}%
      \includegraphics[width=.32\textwidth]{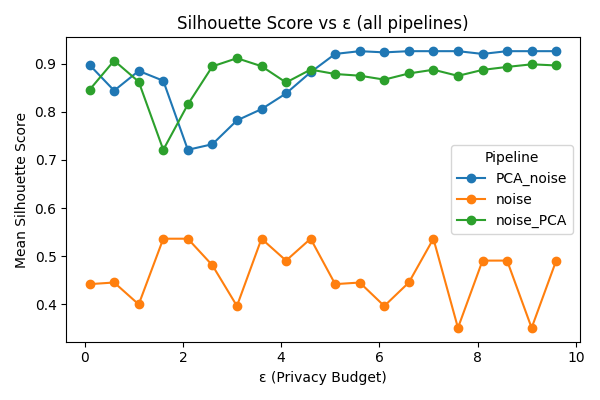} &
      \refstepcounter{panel}\label{fig:CH_epsilon_OpenSNP}%
      \includegraphics[width=.32\textwidth]{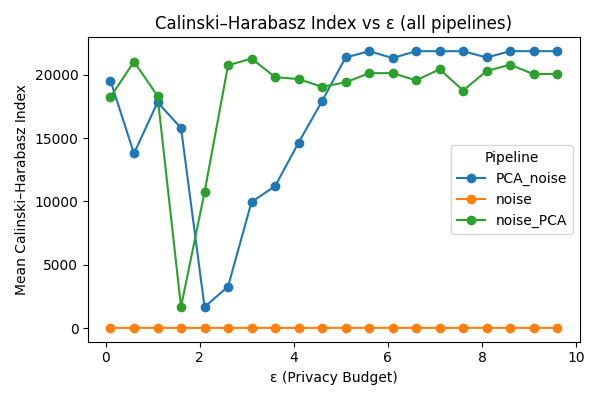} &
      \refstepcounter{panel}\label{fig:DB_epsilon_OpenSNP.png}%
      \includegraphics[width=.32\textwidth]{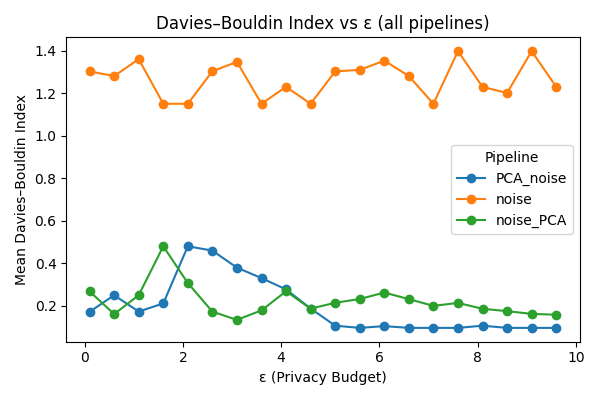} \\
      \small (\ref{fig:Sil_epsilon_OpenSNP}) Silhouette vs.\ $\epsilon$ &
      \small (\ref{fig:CH_epsilon_OpenSNP}) CH Index vs.\ $\epsilon$ &
      \small (\ref{fig:DB_epsilon_OpenSNP.png}) DB Index vs.\ $\epsilon$ \\[0.8ex]

      \refstepcounter{panel}\label{fig:Sil_epsilon_1000genome}%
      \includegraphics[width=.32\textwidth]{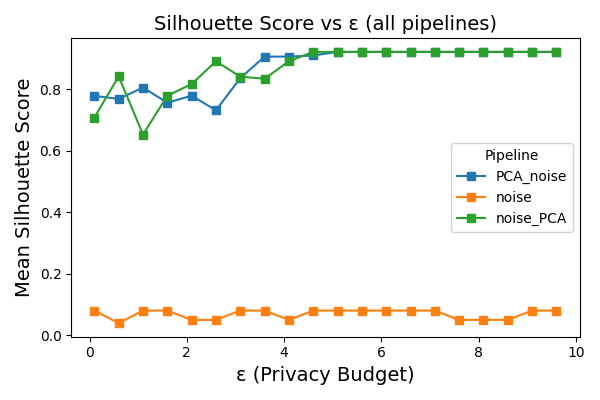} &
      \refstepcounter{panel}\label{fig:CH_epsilon_1000genome}%
      \includegraphics[width=.32\textwidth]{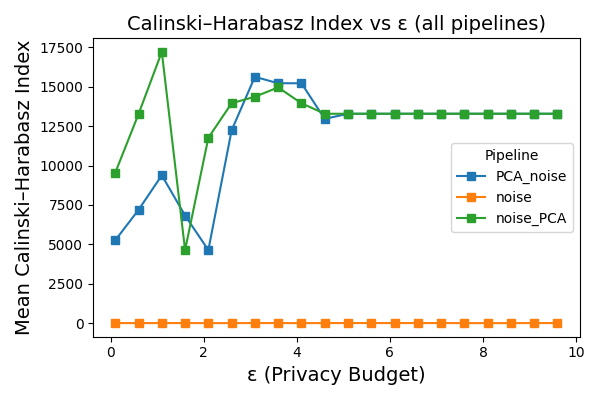} &
      \refstepcounter{panel}\label{fig:DB_epsilon_1000genome}%
      \includegraphics[width=.32\textwidth]{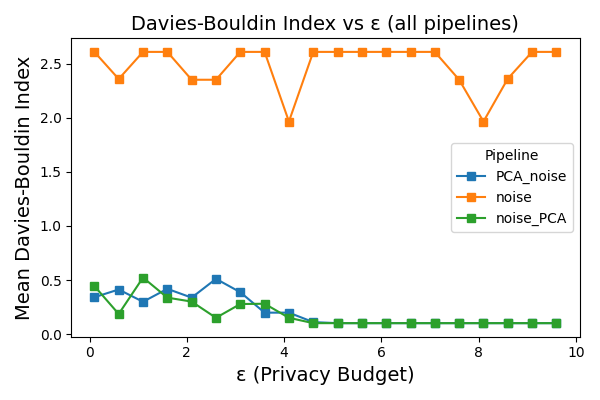} \\
      \small (\ref{fig:Sil_epsilon_1000genome}) Silhouette vs.\ $\epsilon$ &
      \small (\ref{fig:CH_epsilon_1000genome}) CH Index vs.\ $\epsilon$ &
      \small (\ref{fig:DB_epsilon_1000genome}) DB Index vs.\ $\epsilon$ \\

    \end{tabular}%
  }%
}

\caption{Privacy--utility trade-off under the public-data PCA baseline, where the server constructs the PCA model using party shares together with a public dataset. Subfigures (a)--(c) show openSNP results and subfigures (d)--(f) show 1000 Genomes results.}
\label{fig:metrics_epsilon}
\end{figure*}

Figure \ref{fig:metrics_epsilon} reports the same privacy--utility analysis under the public-data PCA baseline, where the server constructs the PCA model using party shares together with a publicly available dataset. The overall trends are consistent with those in Figure \ref{fig:DPCA_metrics_epsilon}: the PCA-based pipelines outperform Noise-Only, and utility improves as $\epsilon$ increases. This comparison shows that the proposed distributed-PCA framework remains close to the public-data baseline while removing the need for an external public dataset.

These empirical trends align with theory as mentioned in section \ref{sec:proposed_work}: by projecting into the top \(r\) principal components before noise injection, PCA$\rightarrow$Noise concentrates the privacy budget into fewer coordinates, reducing the per-coordinate $\ell_1$-sensitivity (and thus the scale of Laplace noise) relative to perturbing all \(d\) raw features \cite{Dwork2014The, chaudhuri2013a}.  Noise$\rightarrow$PCA benefits from a subsequent low-dimensional projection that filters out noise in low-variance directions, yielding intermediate utility \cite{He2025Differentially}, while Noise‐Only disperses noise uniformly across all dimensions and thus suffers the greatest utility loss for any fixed $\epsilon$.

\begin{figure*}[!t]
\centering
\setlength{\tabcolsep}{3pt}
\setcounter{panel}{0}

\alt{figure showing Membership Inference attack powers versus $\epsilon$ for each DP_Pipeline. openSNP (left) and 1000 Genomes (right).}{%
  \mbox{%
    \begin{tabular}{@{}cc@{}}

      \refstepcounter{panel}\label{fig:DPCA_MIA_OpenSNP}%
      \includegraphics[width=.32\textwidth]{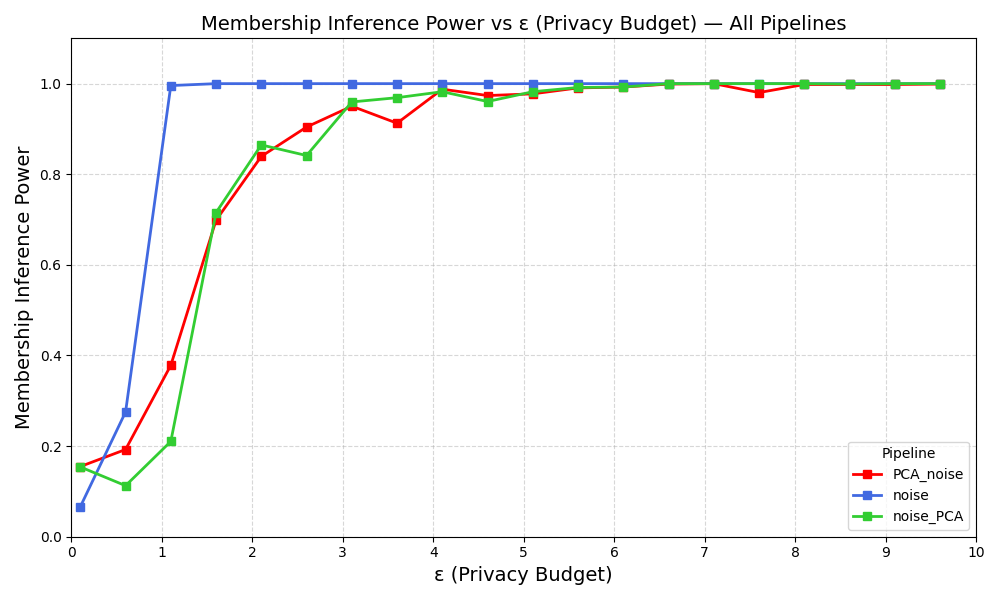} &
      \refstepcounter{panel}\label{fig:DPCA_MIA_1000genome}%
      \includegraphics[width=.32\textwidth]{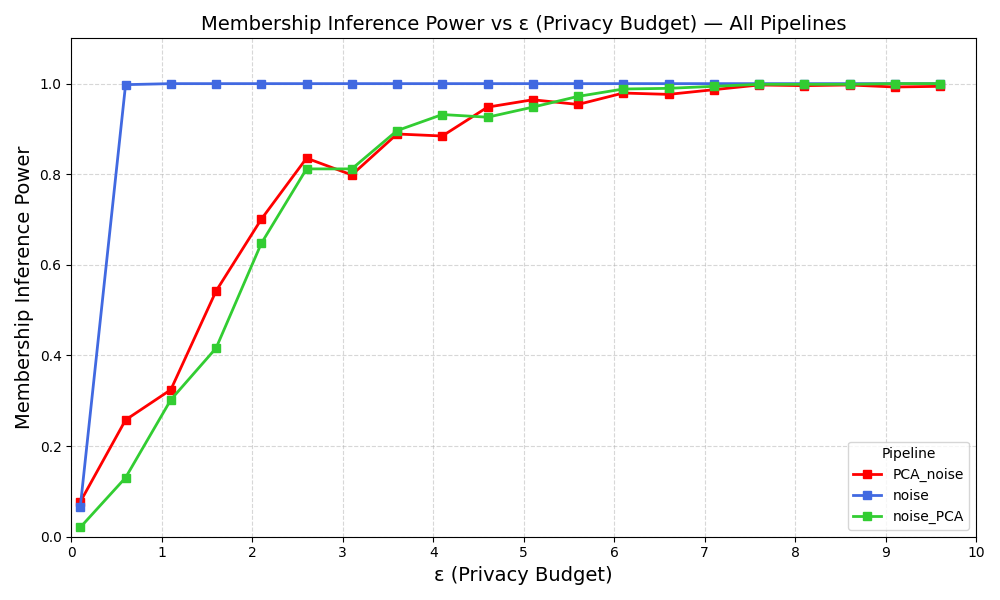} \\
      \small (\ref{fig:DPCA_MIA_OpenSNP}) MIA power vs.\ $\epsilon$  &
      \small (\ref{fig:DPCA_MIA_1000genome}) MIA power vs.\ $\epsilon$  \\[0.8ex]

      \refstepcounter{panel}\label{fig:MIA_OpenSNP}%
      \includegraphics[width=.32\textwidth]{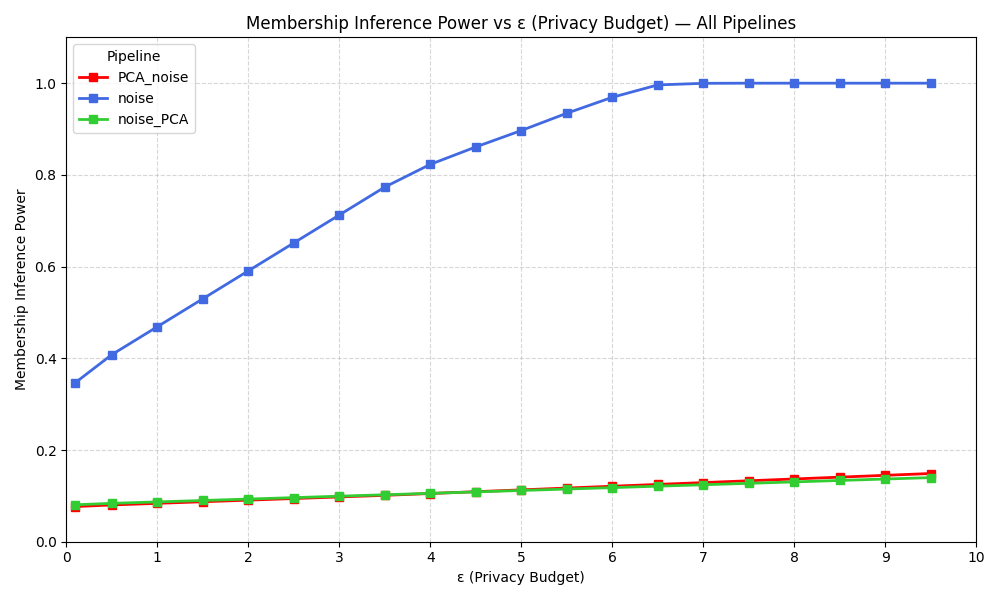} &
      \refstepcounter{panel}\label{fig:MIA_1000genome}%
      \includegraphics[width=.32\textwidth]{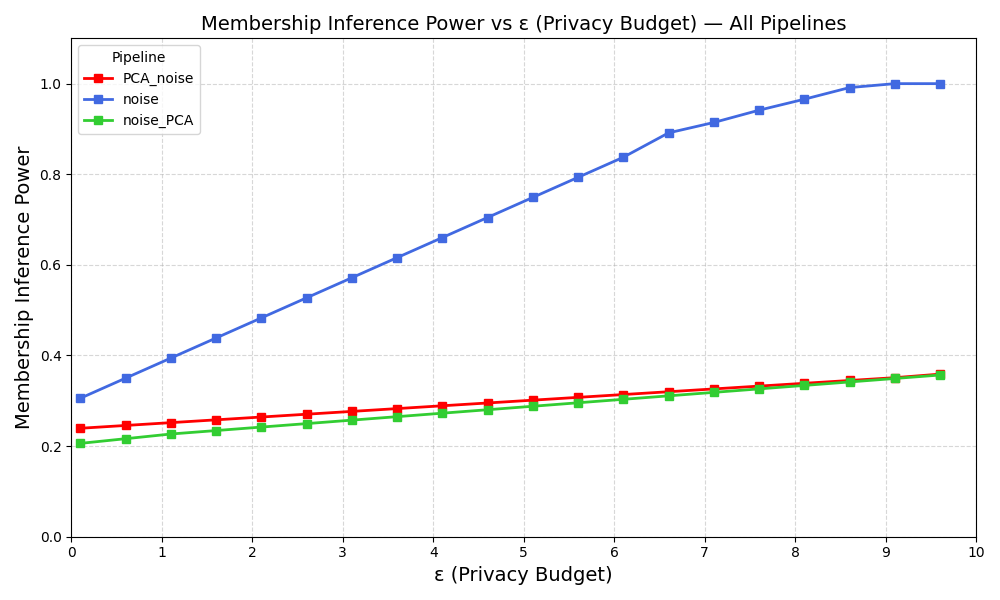} \\
      \small (\ref{fig:MIA_OpenSNP}) MIA power vs.\ $\epsilon$ &
      \small (\ref{fig:MIA_1000genome}) MIA power vs.\ $\epsilon$  \\

    \end{tabular}%
  }%
}

\caption{Membership-inference attack power versus privacy budget.
True-positive rate of a state-of-the-art MIA adversary trained on held-out records under each pipeline. Adding PCA (either before or after noise) sharply suppresses attack power; only Noise-Only curve approaches the maximum. Subfigures (a--b) represent MIA power trends for distributed-PCA frameworks for both openSNP and 1000 Genomes datasets, while (c--d) depict the same for baseline framework (not distributed-PCA). openSNP (left) and 1000 Genomes (right).}
\label{fig:MIA_pipelines}
\end{figure*}

Figure~\ref{fig:MIA_pipelines} complements these utility curves by showing membership‐inference attack (MIA) power versus $\epsilon$. It reports membership-inference attack (MIA) power under the two PCA coordination settings. Subfigures (a)--(b) correspond to the proposed distributed-PCA framework, while subfigures (c)--(d) correspond to the public-data PCA baseline. In the distributed-PCA setting, the PCA-based pipelines exhibit a more gradual increase in attack power as $\epsilon$ grows. This is especially visible in the low-$\epsilon$ regime, where they maintain lower MIA power than Noise-Only. By contrast, Noise-Only approaches high attack power more rapidly, indicating weaker empirical protection against membership inference under the same privacy setting. 
This suggests that dimensionality reduction can improve privacy robustness in practice by concentrating the released representation on the dominant low-dimensional structure rather than exposing a full noisy high-dimensional feature space. The DBSCAN results are consistent with this interpretation: density-based clustering becomes viable after PCA compression, but it remains more sensitive to perturbation than K-Means or HC and deteriorates sharply in the Noise-Only setting.

Taken together, the MIA and utility results indicate that the PCA-based pipelines provide the more favorable overall trade-off in the distributed-PCA framework. PCA$\rightarrow$Noise often delivers the strongest clustering utility, while Noise$\rightarrow$PCA remains a competitive alternative. Noise-Only may still be useful when avoiding PCA is operationally important, but in our experiments it reaches higher attack power more quickly and provides weaker clustering quality unless the privacy budget is relatively large.


\subsection{Privacy–utility and Membership Inference Attack Power}
As $\epsilon$ increases, all pipelines improve in silhouette/CH and reduce DB, with \textit{PCA$\rightarrow$Noise} leading across the range (Figure \ref{fig:DPCA_metrics_epsilon}). MIA power remains lower for PCA-based pipelines: at $\epsilon=3$, \textit{PCA$\rightarrow$Noise} $\approx0.20$ and \textit{Noise$\rightarrow$PCA} $\approx0.35$, while \textit{Noise-Only} is higher (Figure \ref{fig:MIA_pipelines}). 

The privacy trends on 1000 Genomes are consistent with the distributed-PCA setting discussed above. As $\epsilon$ increases, all pipelines improve in silhouette and CH and reduce DB, but the PCA-based pipelines remain clearly stronger than Noise-Only across the privacy range. The MIA curves show the same qualitative behavior: at smaller 
$\epsilon$, the PCA-based pipelines maintain lower attack power, whereas Noise-Only approaches high attack power more rapidly.

\subsection{Runtime and communication}
Table~\ref{tab:running_time} reports the measured runtime and communication cost of the three pipelines on openSNP and 1000 Genomes. Across both datasets, the PCA-based pipelines keep communication in the KB range under the public-PCA baseline and in the sub-MB range under the distributed-PCA setting, whereas Noise-Only requires MB-scale transmission of the full noisy feature matrix.

On openSNP, the aggregate public-PCA communication for the PCA-based pipelines is 5.7 KB, 4.0 KB, and 2.0 KB for $n \in \{710,500,250\}$, respectively, compared with 80.6 MB, 56.8 MB, and 28.4 MB for Noise-Only. Under distributed-PCA, the corresponding aggregate communication is 0.457 MB, 0.456 MB, and 0.455 MB, reflecting the additional one-shot basis-learning overhead together with the upload of noisy PC scores.

A similar trend holds on 1000 Genomes. The aggregate public-PCA communication for the PCA-based pipelines is 3.6 KB, 2.0 KB, and 0.4 KB for $n \in \{450,250,50\}$, respectively, whereas Noise-Only requires 66.6 MB, 37.0 MB, and 7.4 MB. Under distributed-PCA, the total communication is 0.594 MB, 0.593 MB, and 0.592 MB.

Runtime shows the same broad advantage of dimensionality reduction. Noise-Only is consistently the slowest pipeline because the server must evaluate clustering configurations in the original high-dimensional feature space. By contrast, both PCA-based pipelines are substantially faster since clustering is carried out in a low-dimensional representation. The relative ordering between PCA$\rightarrow$Noise and Noise$\rightarrow$PCA depends on the dataset: on openSNP, PCA$\rightarrow$Noise has the lowest total runtime, whereas on 1000 Genomes, Noise$\rightarrow$PCA is faster. Thus, the key conclusion is not that one PCA-based pipeline is always fastest, but that both are markedly more scalable than Noise-Only in terms of both runtime and communication.

A clear sample-size effect is also visible in Table~\ref{tab:running_time}. As the number of collaborative samples decreases, runtime falls for all three pipelines on both datasets, but the reduction is much more pronounced for \textit{Noise-Only} than for the PCA-based pipelines. On openSNP, the total runtime of \textit{Noise-Only} drops from 26.76\,s at $n=710$ to 18.68\,s at $n=500$ and 8.74\,s at $n=250$, while on 1000 Genomes it decreases from 23.68\,s at $n=450$ to 12.51\,s at $n=250$ and 3.06\,s at $n=50$. This stronger dependence on $n$ is expected because \textit{Noise-Only} performs clustering and parameter search directly in the original high-dimensional feature space. By contrast, the PCA-based pipelines remain substantially faster and scale more gently with sample size because the server operates in the reduced $r$-dimensional representation. For example, on openSNP, PCA$\rightarrow$Noise decreases from 2.01\,s to 0.36\,s and 0.27\,s across the same sample sizes, and on 1000 Genomes, Noise$\rightarrow$PCA decreases from 0.76\,s to 0.56\,s and 0.47\,s. Thus, Table~\ref{tab:running_time} shows not only that PCA reduces overall runtime, but also that it makes the framework less sensitive to increases in collaborative sample size.

\FloatBarrier

\begin{table*}[!t]
\caption{Runtime and communication cost of the three DP pipelines on \textit{openSNP} and 1000 Genomes. For each dataset and sample size $n$, we report server-side time (algorithm/parameter search), researcher-side time (local PCA when applicable and noise addition), total runtime, and the per-party communication cost under two PCA settings: a public PCA basis and one-shot distributed-PCA. The PCA-based pipelines communicate only noisy PC scores after PCA learning, whereas \textsc{Noise-Only} transmits the full noisy feature matrix.}
\centering
\begin{adjustbox}{max width=\textwidth}
\begin{tabular}{|c|c|c|c|c|c|c|c|}
\hline
\textbf{Dataset}&\textbf{Pipeline}&\textbf{n}& \textbf{Server-side time}& \textbf{Researcher-side time}& \textbf{Total running time} & \textbf{Public PCA bandwidth} & \textbf{Distributed-PCA bandwidth}\\ 
\hline
     &&    710 & 1.71 S & 0.3 S & 2.01 S & 5.7 KB & 0.457 MB\\
     openSNP &PCA$\rightarrow$noise  &  500&  0.12 S & 0.24 S & 0.36 S& 4.0 KB & 0.456 MB\\
     &&    250    &   0.07 S & 0.2 S   & 0.27 S & 2.0 KB & 0.455 MB \\
     \hline
     &&    710&  1.40 S &  12.5 S &  13.9 S & 5.7 KB& 0.457 MB \\
    openSNP& noise$\rightarrow$PCA &   500 & 0.12 S  & 10.75 S &  10.87 S & 4.0 KB & 0.456 MB \\
     &&   250 & 0.08 S  & 5.1 S & 5.18 S & 2.0 KB & 0.455 MB \\
     \hline
     &&  710  & 14.43 S &  12.33 S &  26.76 S & 80.6 MB & 0.457 MB\\
     openSNP &Noise-Only&     500 &9.60 S  & 9.08 S & 18.68 S & 56.8 MB & 0.456 MB\\
     &&  250 & 3.68 S & 5.06 S &  8.74 S & 28.4 MB & 0.455 MB \\
     \hline
     &&    450 & 0.46 S & 14.21 S& 14.67 S & 3.6 KB & 0.594 MB\\
     1000Genomes&PCA$\rightarrow$noise  &  250&  0.35 S & 6.48 S & 6.83 S & 2.0 KB & 0.593 MB\\
     &&    50    &   0.27 S & 1.09 S  & 1.36 S & 0.4 KB & 0.592 MB\\
     \hline
     &&    450&  0.43 S &  0.33 S &  0.76 S & 3.6 KB & 0.594 MB \\
     1000Genomes& noise$\rightarrow$PCA &   250 & 0.29 S  & 0.27 S &  0.56 S & 2.0 KB & 0.593 MB \\
     &&   50 & 0.26 S & 0.21 S & 0.47 S & 0.4 KB & 0.592 MB \\
     \hline
     &&  450  & 10.49 S &  13.19 S&  23.68 S & 66.6 MB  & 0.594 MB\\
    1000Genomes& Noise-only&     250 & 5.53 S  & 6.98 S & 12.51 S & 37.0 MB & 0.593 MB \\
     &&  50 & 1.95 S  & 1.11 S &  3.06 S & 7.4 MB & 0.592 MB\\
     \hline
\end{tabular}
\end{adjustbox}
\label{tab:running_time}
\end{table*}

\subsection{Computational Complexity} \label{sec:complexity}
We analyze both computational and communication costs in the collaborative setting. Let party $i$ hold $n_i$ samples with $d$ SNPs; let $N=\sum_i n_i$ be the total number of collaborative samples, let $r \ll d$ denote the number of retained principal components, and let $M$ denote the number of candidate algorithm--parameter settings evaluated by the server.

\textbf{Party-side cost and communication.}
For the two PCA-based pipelines, once a shared PCA basis is available, each party projects its local $n_i \times d$ matrix onto a $d \times r$ basis and perturbs only the resulting $n_i \times r$ score matrix. The dominant local work is therefore the projection step, which costs $O(n_i d r)$, while the local noise-injection step costs $O(n_i r)$. The clustering-stage communication is correspondingly reduced to $O(n_i r)$ values per party, instead of transmitting the full $n_i \times d$ matrix.

In the \textit{public-PCA} baseline, the shared basis is constructed once on public data and broadcast to the parties, so each party only uploads its noisy PC scores during collaborative clustering. In the \textit{distributed-PCA} setting, each party also computes a local top-$r$ basis and sends this one-time low-rank summary to the server, which aggregates the local eigenspaces and returns the final shared basis. Thus, the total communication in the distributed-PCA setting is more accurately described as a one-time $O(d r)$ basis-exchange cost plus the usual $O(n_i r)$ clustering-stage upload. This distinction matches Table~\ref{tab:running_time}, where the public-PCA bandwidth is in the KB range and the distributed-PCA bandwidth includes the additional one-shot basis-learning overhead.

For \textit{Noise-Only}, no dimensionality reduction is performed. Each party perturbs all $d$ features and uploads the full $n_i \times d$ noisy matrix, giving local work of $O(n_i d)$ and communication of $O(n_i d)$ values per party. As reflected in Table~\ref{tab:running_time}, this full-dimensional release leads to substantially higher bandwidth than the PCA-based pipelines.

\textbf{Server-side clustering and parameter search.}
After collecting the LDP-protected shares, the server evaluates multiple clustering algorithms over a grid of hyperparameter settings. Let $d' = r$ for the PCA-based pipelines and $d' = d$ for Noise-Only. For a single candidate configuration, the dominant server-side costs are:

\begin{itemize}
    \item \textbf{K-Means:} $O(N K d' T)$, where $K$ is the number of clusters and $T$ is the number of Lloyd iterations.
    \item \textbf{Agglomerative Clustering (HC):} typically $O(N^2 d')$ to form pairwise distances, with additional linkage-update cost depending on the implementation.
    \item \textbf{GMM:} roughly $O(N K d' T_{\mathrm{EM}})$ per EM run, plus covariance-related overhead that grows with $d'$.
    \item \textbf{DBSCAN:} with an efficient spatial index, neighborhood queries are typically near $O(N \log N)$ in low dimension, but the worst case is $O(N^2)$; in practice, DBSCAN becomes substantially more expensive and less stable as the ambient dimension grows because neighborhood search degrades toward all-pairs distance computation.
\end{itemize}

Since the server evaluates $M$ algorithm--parameter settings, the total search cost scales with the sum of these per-configuration costs over the tested grid. The key implication is that all four algorithms benefit from replacing $d$ with $r$ in the PCA-based pipelines, but DBSCAN benefits especially from this reduction because density-based neighborhood search is much more practical in a compact low-dimensional representation than in the original high-dimensional SNP space.

This interpretation is consistent with both Table~\ref{tab:combined_clustering} and Table~\ref{tab:running_time}. In Table~\ref{tab:combined_clustering}, DBSCAN returns meaningful partitions for the PCA-based pipelines but is weaker than K-Means and HC, and it fails under Noise-Only on both datasets. In Table~\ref{tab:running_time}, the server-side search time is far lower for the PCA-based pipelines than for Noise-Only, because all candidate algorithms---including DBSCAN---are evaluated in $r$ dimensions rather than $d$. Therefore, the main computational advantage of PCA in our framework is not only lower communication, but also the fact that it makes the entire server-side model-selection procedure substantially more scalable.

\FloatBarrier
\section{Conclusions}\label{sec:conclusion}
We presented a privacy-preserving framework for collaborative population stratification workflows, whose purpose is to recommend a strong clustering algorithm–parameter pair from privacy-protected shares. The core idea is to allow multiple parties to contribute only sanitized information while enabling the server to evaluate candidate clustering algorithms and hyperparameter settings directly on LDP-protected data. This makes it possible to recommend a strong clustering configuration without incurring any privacy loss beyond the local $\epsilon$-LDP mechanisms.

The experiments on openSNP and 1000 Genomes support three main conclusions. First, the server-side recommendation layer is effective: across the evaluated settings, it consistently identifies one of the top-performing algorithm--parameter configurations from sanitized data, even when several methods are nearly tied. Second, the PCA-based pipelines, \textit{PCA$\rightarrow$Noise} and \textit{Noise$\rightarrow$PCA}, provide substantially stronger clustering utility than \textit{Noise-Only}. This confirms that performing collaborative clustering in a low-dimensional representation is much more effective than perturbing and clustering directly in the original high-dimensional SNP space. Third, the systems results show that the PCA-based pipelines drastically reduce both communication cost and server runtime, since they exchange only low-dimensional score matrices rather than full noisy genotype matrices.

A further finding is that the three pipelines expose different trade-offs rather than a single absolute ordering. Among them, \textit{PCA$\rightarrow$Noise} often yields the strongest clustering utility, while \textit{Noise$\rightarrow$PCA} remains a competitive alternative and may be computationally preferable on some datasets. \textit{Noise-Only} avoids the projection step but incurs the largest communication burden and the highest server-side clustering cost, and its clustering quality is consistently weaker in our experiments. In addition, under the distributed-PCA framework, the PCA-based pipelines exhibit lower membership-inference attack power in the low-$\epsilon$ regime, whereas \textit{Noise-Only} reaches high attack power more rapidly. Thus, distributed-PCA processing improves not only scalability and utility, but also empirical privacy robustness at stronger privacy settings.

The journal version also broadens the algorithmic scope of the framework by incorporating DBSCAN into the server-side search space in addition to K-Means, GMM, and HC. In the reported experiments, DBSCAN produces meaningful partitions on the PCA-based pipelines but remains slightly below the best K-Means/HC configurations, and it becomes unstable in the Noise-Only setting. Its inclusion therefore improves the generality of the selection layer even though the strongest results in our datasets are still obtained by the PCA-based centroid- and hierarchy-based methods.

Overall, the evaluation supports three conclusions. First, the server-side recommendation layer can identify strong algorithm--parameter settings directly from perturbed data. Second, under the proposed distributed-PCA framework (Figure \ref{fig:DPCA_metrics_epsilon}), the PCA-based pipelines provide substantially better clustering quality than Noise-Only across both datasets. Third, comparison with the public-data PCA baseline (Figure \ref{fig:metrics_epsilon}) shows that the distributed-PCA framework preserves similar privacy--utility behavior while avoiding reliance on a public dataset. In addition, the distributed-PCA MIA results in Figure \ref{fig:MIA_pipelines}(a)--(b) show that the PCA-based pipelines maintain lower attack power than Noise-Only in the low-$\epsilon$ regime.
\balance
\bibliographystyle{IEEEtran}
\bibliography{References/ref}

\end{document}